# Spin Hall Switching of the Magnetization in Ta/TbFeCo Structures with Bulk Perpendicular Anisotropy

Zhengyang Zhao, Mahdi Jamali, Angeline Klemm, and Jian-Ping Wang [*]

Department of Electrical and Computer Engineering, University of Minnesota, 4-174 200 Union Street SE, Minneapolis, MN 55455, USA

Spin-orbit torques are studied in Ta/TbFeCo patterned structures with a bulk perpendicular magnetic anisotropy (bulk-PMA) for the first time. The current-induced magnetization switching is investigated in the presence of a perpendicular, longitudinal, or transverse field. In order to rule out Joule heating effect, switching of the magnetization is also demonstrated using current pulses. It is found that the anti-damping torque correlated with spin Hall effect is very strong, and a spin Hall angle of about 0.12 is obtained. The field-like torque related with Rashba effect is negligible in this structure suggesting that the interface play a significant role in Rashba-like torque.

*Corresponding author. Tel: (612) 625-9509. E-mail:   jpwang@umn.edu

Recently current-induced spin-orbit torques have been intensively studied in magnetic ultra-thin films due to its potential for memory and logic devices.[1–3] Researchers have reported that the magnetization of ferromagnetic structures with interfacial perpendicular magnetic anisotropy (I-PMA) including Pt/Co/Oxide,[4–7] Pt/Co/Pt,[8,9] Ta/CoFeB/MgO,[10–14] and Ta/CoFe/MgO[15] can be switched by in-plane current-induced spin-orbit torques (SOT), namely, the spin Hall (SH) and the Rashba effects. It provides an efficient way of generating spin current that promises to significantly reduce the current required for switching PMA nanomagnets. However, the interfacial-PMA is usually not strong enough to preserve the thermal stability of nanomagnets for applications in memory and logic devices at sub-20 nm,[16] although different methods have been developed to enhance the interfacial-PMA.[17,18] On the other hand, bulk-PMA materials, including $L1_0$-ordered alloys (FePt, FePd, CoPt, etc.),[19,20] Heusler alloys,[21,22] and transition metal-rare earth alloys,[23] exhibit a much larger $K_u$ (~$10^7$ erg/cm$^3$) and higher thermal stabilities; however, the manipulation of magnetization with bulk-PMA using SOT is rarely studied so far.[24]

The utilization of bulk-PMA materials in SOT/spin Hall systems also has unique advantages for understanding the underlying physics of SOT. In SOT systems, the contributions of the interfacial effect (Rashba) and the bulk effect (spin Hall) are controversial.[4,25,26] In magnetic materials with bulk-PMA, magnetic anisotropy originates from the bulk rather than the interface; hence, it is a suitable system for the investigation of SOT. On the contrary to the interfacial-PMA structures, the materials with bulk-PMA are much more robust to thickness or interface variation. This makes it possible to do a comparative study with the same bulk-PMA material and different heavy metals. Additionally, it also allows the study of changing the thickness of the magnetic layer, which is restricted when dealing with interfacial-PMA structures.[10]

In this letter, we first demonstrate the spin Hall effect induced switching of perpendicular magnetization with a bulk-PMA. Ta/TbFeCo structure is employed to study the magnetization switching in materials with strong intrinsic perpendicular anisotropy. The current-induced magnetization switching is investigated in the presence of a perpendicular field, a longitudinal field, or a transverse field. We determine the strength of anti-damping torque (or spin Hall efficiency) is in accordance with that in the interfacial-PMA system Ta/CoFeB/MgO reported previously, while the field-like torque (Rashba torque) is negligible.

The film stack consists of, from the substrate, Ta(5)/Tb$_{20}$Fe$_{64}$Co$_{16}$(1.8)/MgO(2)/Ta(4) (thickness in nm). The films are deposited on thermally oxidized silicon wafers by DC and RF magnetron sputtering at room temperature. A ternary alloy target is used for the deposition of the TbFeCo layer. The composition of the TbFeCo is confirmed using RBS and XPS measurements. The as-deposited films demonstrate a strong intrinsic perpendicular anisotropy[27–29] as shown in Fig. 1(a). The 1.8 nm-thick TbFeCo film shows PMA with a square hysteresis loop in the out-of-plane direction with a coercivity field $H_C$ = 70 Oe. The in-plane hysteresis loop is saturated at about 10 kOe indicating a strong perpendicular anisotropy field of 1.5 T. With the thickness of TbFeCo increasing, the out-of-plane loops persist good squareness and sharp switching, as can be seen from the inset of Fig. 1(a), indicating the PMA in TbFeCo originates from the bulk rather than induced by the interface. The coercivity field increases upon increasing the TbFeCo thickness, which is a feature of bulk-PMA materials. We also compared the hysteresis loops of 1.8 nm TbFeCo deposited on different giant spin Hall metals (Ta and W), and both systems exhibit similar perpendicular properties, which implies the robust behavior of the bulk-PMA in TbFeCo. Fig. 1(b) shows the experimental setup and device schematic. The film stack is patterned into Hall bars with a width of 10 μm and length of 65 μm using optical lithography and Ar-ion etching. A DC current is injected into the bar along the longitudinal direction (y direction), and the voltage is detected by a nano-voltmeter in the transverse direction (x direction). The anomalous Hall resistance, $R_H$, which is proportional to the perpendicular component of the magnetization, $M_Z$, is measured to determine the magnetization direction of the TbFeCo layer. The polarity of $R_H$ depends on the orientation of magnetization, and positive $R_H$ corresponds to $M_z > 0$.

We started by investigating the variation of the hysteresis loop while gradually changing the current, and the results are shown in Fig. 2(a). The anomalous Hall resistance is measured as a function of perpendicular magnetic field $H_z$ for different input currents. When the current flowing through the device is very small (0.5 mA), we observe a square hysteresis loop with coercivity of 240 Oe and Hall resistance with the amplitude of 4.0 Ω. The coercivity is larger than that of unpatterned thin film in Fig. 1(a) because of the incoherent switching in unpatterned film.[30] Once the current is increased to 8 mA, the coercivity reduces dramatically to 50 Oe, and the amplitude of $R_H$ also decreases to 3.65 Ω. The reduction of both the coercivity and the Hall resistance results from the increase of current-induced torque related with the spin Hall effect, or the anti-damping torque, expressed as $\boldsymbol{\tau}_\| = \tau_\|^0 \mathbf{m} \times (\mathbf{m} \times \hat{\boldsymbol{\sigma}})$, where $\hat{\boldsymbol{\sigma}}$ denotes the spin polarization unit vector, and the magnitude of the torque is $\tau_\|^0 = \frac{\hbar}{2eM_s t}|J_S|$, where $J_S$ represents the spin current. For a positive charge current flowing through the Hall bar (along +y), $\boldsymbol{\tau}_\|$ is along +x. This torque, on one hand, tilts the magnetization from the z direction towards the y direction in the y-z plane and decreases the energy barrier for switching. As shown in the phase diagram in Fig. 2(a), the perpendicular coercivity can be modulated from 240 Oe to 10 Oe by varying the current from 0.5 mA (corresponding to a current density of $0.25 \times 10^6$ A/cm² in the Ta layer) to 12 mA ($3.75 \times 10^6$ A/cm² in the Ta layer).

The influence of the SHE on the magnetization can also be studied by sweeping the current in the presence of various in-plane fields along the y direction, as seen in Fig. 2(b). When the current is applied parallel with $H_y$, the direction of the effective field $\mathbf{H}_\| = \tau_\|^0 (\mathbf{m} \times \hat{\boldsymbol{\sigma}})$, which is corresponding to $\boldsymbol{\tau}_\|$, is oriented parallel with $H_y$ for $M_z < 0$ and anti-parallel with $H_y$ for $M_z > 0$;

therefore the upward magnetization is favored. Similarly, when the current is applied anti-parallel with the $H_y$ field, downward magnetization is favored. Thus by sweeping the DC current, the switching of the magnetization can be observed, and the polarity of the switching loop is dependent on the direction of $H_y$ (inset of Fig. 2(b)) and is in agreement with the observations in the Ta/CoFeB/MgO system in the previous study.[12] The value of switching current decreases from 12 mA to 5 mA as the in-plane field increases from 50 Oe to 600 Oe, in agreement with the scenario based on the current induced torque $\tau_\parallel$ related with the spin Hall effect.[4]

In order to rule out Joule heating effects, we perform the experiment with pulse current, as exhibited in Fig. 3 (a) and (b). The current pulse is injected in the +y direction with an amplitude of 10 mA and duration of 1 ms. The Hall resistance is measured 20 μs after rise time of the current pulse. A perpendicular field is applied with the field direction reversed back and forth after each current pulse. When the field strength is 50 Oe, the Hall resistance switches back and forth synchronously, varying between +3.6 Ω and -3.6 Ω (Fig. 3(a)). It indicates that the magnetization switches between upward state and downward state completely. The osilation of magnetization is captured by the magneto-optical Kerr effect (MOKE) images, as shown in Fig. 3(c), where the two MOKE images of the Hall bar with different brighness correspond to $M_z < 0$ ($R_H = -3.6$ Ω) and $M_z > 0$ ($R_H = +3.6$ Ω) repectively. On the other hand, when the strength of the applied field decreases to 20 Oe for the same current pulse, the Hall resistance only varies between +3.6 Ω and approximate +2.5 Ω (Fig. 3(b)). Such small variation in $R_H$ indicates the magnetization can't be completed switched with 20 Oe field, and barely small reverse domains are nucleated. The results in Fig. 3 (a) and (b) are consistent with the switching diagram in Fig. 2 (a): for the current of 10 mA, $H_z$ of 20 Oe is located near the boundary of the switching diagram and thereby barely enough for domain wall nucleation, while $H_z$ of 50 Oe is located in the reversal area in the diagram and thereby forming a completed switching of magnetization.

We obtain more insight by sweeping the current in the presence of a transverse in-plane field, $H_x$, instead of the longitudinal field, $H_y$, as can be seen in Fig. 4. When $H_x$ is zero or smaller (Fig. 4 (a)), the contour of Hall resistance in the shape of an arch is obtained with the current varying from -20 mA to +20 mA or vice versa. The gradual drop of $R_H$ from about 4 Ω at $I = 0$ mA to less than 2 Ω at $I = 20$ mA reflects the magnetization tilting due to the current induced torque. No switching is obtained here, because the transverse field $H_x$, which is vertical to the effective field $\mathbf{H}_\parallel$ of spin Hall effect, can't break the symmetry. However, when $H_x$ becomes larger (Fig. 4 (b) and (c)), the contour of $R_H$ changes noncontinuously with two bumps appearing. This could be a result from the joint effect of SHE torque, transverse field, and Dzyaloshinskii-Moriya interaction (DMI). For example when $H_x = 500$ Oe, as the current varies from negative to zero, $R_H$ increases gradually but is still much less than 4Ω at $I = 0$; this means the magnetization is still tilted significantly from the $z$-axis. The external field itself is not strong enough to lead to this state. Instead, DMI could be a reason for such an intermediate state by generating stable helical magnetization patterns.[31] With the current increasing further to positive values, $R_H$ jumps up abruptly to nearly 4Ω (at $I = 7$ mA) and then jumps back (at $I = 13$ mA), due to the competition of DMI and the SHE. When current is swept back from positive to negative, $R_H$ follows a similar variation. This "half switching" phenomenon is not previously reported in Ta/CoFeB/MgO or another interfacial-PMA system. Whether it is or isn't a unique feature of bulk-PMA materials still requires further study.

Finally, we quantitatively evaluated the strength of current-induced torques. In this case, we apply an in-plane field and compare field sweeps for the same magnitude of current, positive and negative ($I = \pm 4$ mA in Fig. 5). Fig. 5(a) shows the result for when the field is applied along the longitudinal direction. In this case, the difference in the Hall resistances for positive current and negative current is attributed to the anti-damping torque related with spin Hall effect, following the equation $\Delta H_y = \tau_\parallel^0 / \sin\theta$, where $\theta$ denotes the angle between the magnetization and $x$-$y$ plane.[4]

We calculate $\tau_\parallel^0 / I = 1.5$ mT/mA and spin Hall angle is estimated to be about $J_S / J_C = 0.12$. This value is similar with that previously reported in Ta/CoFeB/MgO system.[1] Then we repeat this measurement with the field applied along the transverse direction as shown in Fig. 5(b). In this case, there's no discernable difference between the curves with $I = +4$ mA and $I = -4$ mA indicating the field-like torque $\tau_\perp$ related with the Rashba effect is negligible in the Ta/TbFeCo system. This is predictable because the Rashba effect is an interfacial effect originating from the ultrathin asymmetric sandwiched structure, which isn't the case for our bulk-PMA system.

In summary, bulk-PMA structures can overcome the limitations of interfacial-PMA structures in the SOT study, and therefore are promising candidates in both the physical understanding the industrial application of SOT or spin Hall effect. We studied the spin Hall effect in Ta/TbFeCo structures with bulk-PMA for the first time. The current-induced magnetization switching is studied in the presence of a longitudinal field while the "half-switching" phenomenon is obtained in the presence of a transverse field. The strength of anti-damping torque (or spin Hall efficiency) is in accordance with that in Ta/CoFeB/MgO systems reported previously, and the field-like torque (Rashba torque) is negligible in Ta/TbFeCo system.

Acknowledgement

This work was partially supported by the Center for Spintronic Materials, Interfaces and Novel Architectures (C-SPIN), one of six SRC STARnet Centers, sponsored by MARCO and DARPA and the National Science Foundation Nanoelectronics Beyond 2020 (Grant No.NSF NEB **1124831**).

Figure Captions

FIG. 1. (a) The in-plane and out-of-plane hysteresis loop of Ta(5)/TbFeCo(1.8)/MgO(2) multi-layers. Inset: the out-of-plane hysteresis loops for Ta(5)/TbFeCo(t)/MgO(2) multilayers, where $t = 1.8$, 2.2, and 2.5 nm, respectively. (b) Schematic illustration of Hall bar device and setup of the experiment

FIG. 2. (a) Out-of-plane field dependence of $R_H$ with $I = 0.5$ mA and 8mA respectively (inset), and switching phase diagram where the coercivity varies as a function of current. (b) Current dependence of $R_H$ in the presence of longitudinal field $H_y = 390$ Oe and -390 Oe, respectively (inset), and switching phase diagram where the switching current $I_S$ varies as a function of $H_y$.

FIG. 3. (a) and (b) $R_H$ variation with injection of a sequence of current pulses of $I_p = 10$ mA and 1 ms long; the perpendicular field is applied and field direction reversed after each current pulse. The amplitude of the applied field is $H_z = 50$ Oe for (a) and $H_z = 30$ Oe for (b). (c) Magneto-optical Kerr image of the Hall bar device with the magnetization direction pointing upward (left) and downward (right), respectively.

FIG. 4. $R_H$ as a function of DC current when the transverse field $H_x$ is present. (a) $H_x = 120$ Oe, (b) $H_x = 200$ Oe and (c) $H_x = 500$ Oe.

FIG. 5. $R_H$ as a function of the (a) longitudinal field or (b) transverse field, for $I = \pm 4$ mA respectively. Inset of (a): the difference in the field for +2 mA and -2 mA when the Hall resistance is the same value.

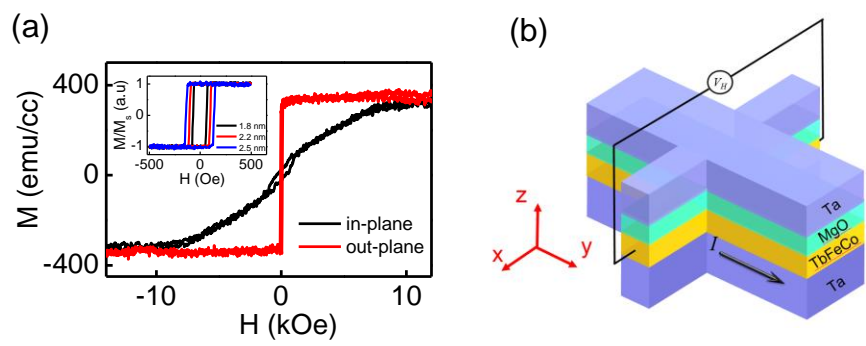

Figure 1.

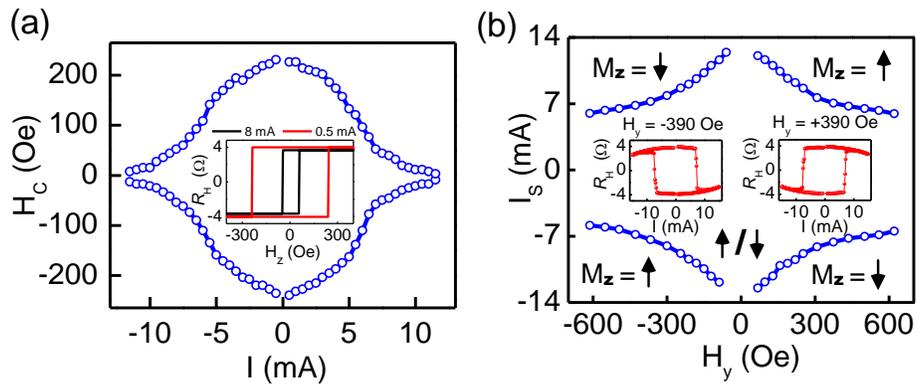

Figure 2.

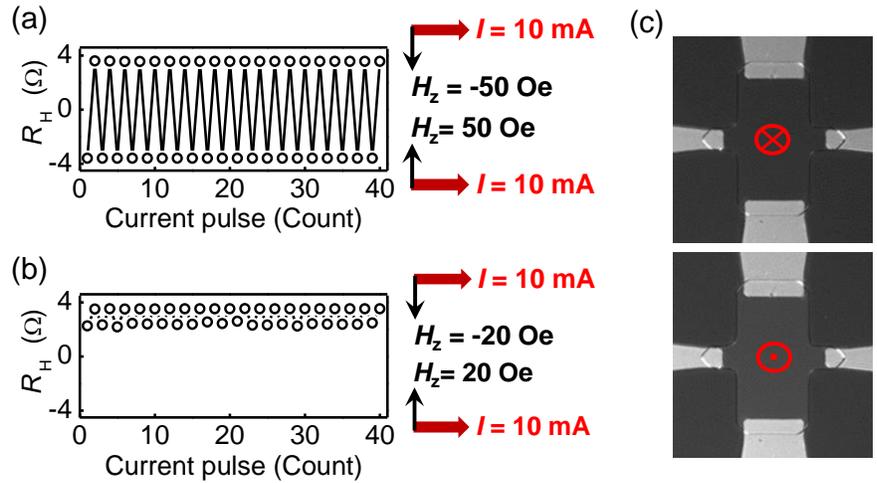

Figure 3.

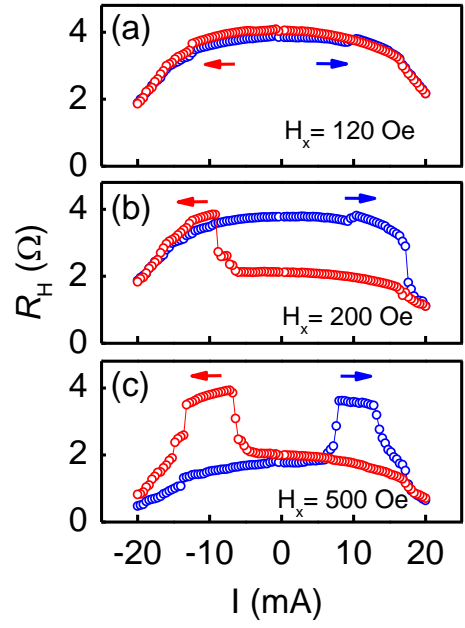

Figure 4.

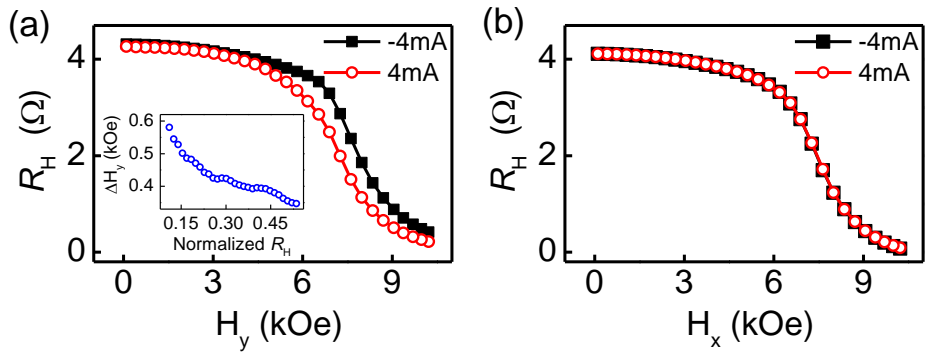

Figure 5.